\documentclass[12pt,fleqn]{cernart}
\usepackage{epsfig}
\usepackage{times}
\begin{document}

\renewcommand{\topfraction}{0.99}
%erlaubt, dass ein Bild am oberen Rand 99% der Seite einnehmen darf.
\renewcommand{\bottomfraction}{0.99}
%erlaubt, dass ein Bild am unteren Rand 99% der Seite einnehmen darf.
\renewcommand{\textfraction}{0.01}
%erlaubt, dass Text >= 1% der Seite einnehmen muss.
\setcounter{totalnumber}{6}
%erlaubt maximal 6 Bilder (bewegliche Objekte) pro Seite

\begin{titlepage}
%\begin{frontmatter}
%\docnum{DRAFT 1.2}
%\docnum{CERN/SPSC 2002-038}
%\docnum{CERN/SPSC/P264 Add.~11}
\vspace{1.0cm}
\date{12 January 2004}

%\vspace{2.0cm}
%\begin{center}
%{\large ADDENDUM--11 to PROPOSAL CERN/SPSC/P264}
%\end{center}

\vspace{2.0cm}
\title{\large{
Are there S=-2 Pentaquarks?
}}

\vspace{0.5cm}
\begin{center}
{\large H.G.~Fischer and S.~Wenig }
\vspace{4mm}  \\
CERN  \\
%\vspace{2mm}
Geneva, Switzerland
\end{center}

\vspace{3.0cm}
\begin{abstract}
Recent evidence for pentaquark baryons in the channels $\Xi^-\pi^-$, $\Xi^-\pi^+$ and their 
anti-particles claimed by the NA49 collaboration is critically confronted with the
vast amount of existing data on $\Xi$ spectroscopy which was accumulated over the past decades. 
It is shown that the claim is at least partially inconsistent with these data. 
In addition two further exotic channels of the pentaquark type available in the NA49 data are
investigated. 
It is argued that this study leads to internal inconsistency with the purported signals.

\end{abstract}

%\end{frontmatter}
%\cleardoublepage

%\newpage

%\begin{Authlist}
%\vspace{2.0cm}
%\noindent
%\end{Authlist}

\end{titlepage}

\newpage

\section{Introduction
}
Over the past few months a number of observations of an exotic baryonic
state in the NK channel, the $\Theta^+$(1540), have been claimed
[1--6].
A common feature of all these claims is a relatively low event statistics
with between 20 and 100 entries in the peaks, a signal to background
ratio ranging from 1:1 to 1:3, a statistical significance in the
region from 3 to 5 standard deviations and, whenever cross section
estimations have been made available, a rather large $\Theta^+$(1540)/$\Lambda$(1520)
ratio of about 1/2 to 1/3.

This evolution has prompted the search for other members of the
corresponding minimal pentaquark anti-decuplet, especially 
the $S=-2$ isospin-quadruplet. In fact, the NA49 collaboration has claimed the
observation of signals in the $\Xi^-\pi^-$ ($I_3=-3/2$) and $\Xi^-\pi^+$ ($I_3=+1/2$) 
combinations and their anti-states \cite{pen}. As far as numbers of entries, 
signal to background ratios and statistical significance are concerned, these
claims show surprising similarities to the features mentioned above for 
the $\Theta^+$ state.

If the claims for the $\Theta^+(1540)$ might benefit, at least for the nK$^+$
decay, from the absence of preceding high statistics experiments, this
is however not true for the $S=-2$ quadruplet. Several decades of experimental
work in $\Xi$ spectroscopy have yielded a large thesaurus of data which
should be carefully consulted before making any statements about new
baryonic states from low statistics data.

In this note we attempt, in Section~2, a comparative study of
existing data on $\Xi$ spectroscopy. In Section~3, we address the
minimal internal consistency to be required between the different
pentaquark states accessible in the NA49 data. Finally, in Section~4 we recall 
some facts concerning past experience with low statistics spectroscopy.

\section{$\Xi$-Spectroscopy
}
With the availability of K$^-$ beams at different accelerators a vivid
activity concerning the study of $\Xi$ hyperons, their excited states
as well as exotic mass combinations containing the $\Xi$, started in
the mid-1960's. 
When these studies were finally abandoned in the mid-1980's, the available 
data spanned a wide range of beam momenta from about 2~GeV/c to 16~GeV/c 
both on proton and deuteron targets. 
The statistical significance of these data also covered a wide range, 
from a few hundred measured $\Xi^-$ in the early experiments up to large 
statistics samples containing about 10$^4$ $\Xi^-$.

In the SPS energy range, experiments with hyperon beams, especially
using $\Sigma^-$ and $\Xi^-$, cover the range from 100 to 350 GeV/c
beam momentum. Some results from these experiments which have been terminated
more than a decade ago, have been published only recently.
These results benefit from an unprecedented event statistics reaching up
into the region of more than $2\cdot10^5$ detected $\Xi^-$ and 
$6\cdot10^4$~$\Xi^{*0}$(1530) hyperons.

A straight-forward, non-exhaustive literature search [8--31] shows 
about 30 different experiments covering the two main classes mentioned 
above.

After some initial claims of exited $\Xi$ states in different mass ranges,
the advent of high statistics data revealed a puzzle that persists
even today, namely the difficulty to discern any structure in the $\Xi\pi$ 
mass spectra above the well-known $\Xi^*$(1530) resonance. 
This difficulty concerns exotic as well as  non-exotic decay channels in
a situation where at least in the non-exotic sector a rich spectroscopy
in analogy to the dense coverage of N$^*$ states would be expected.

Given this situation, any experiment attempting the study of $\Xi$
spectroscopy today, especially in p+p interactions with their
grave penalty both from initial strangeness and from initial charge content,
should carefully confront these rich data samples. In particular
it should be made sure that the new data do not fail to come up
to the quality of the earlier work.

In the following we attempt to summarize this experimental situation,
essentially drawing on published results.

\subsection{The $\Xi^-\pi^+$ and $\Xi^-\pi^0$ Channels
}
\begin{figure}
\centering
\epsfig{file=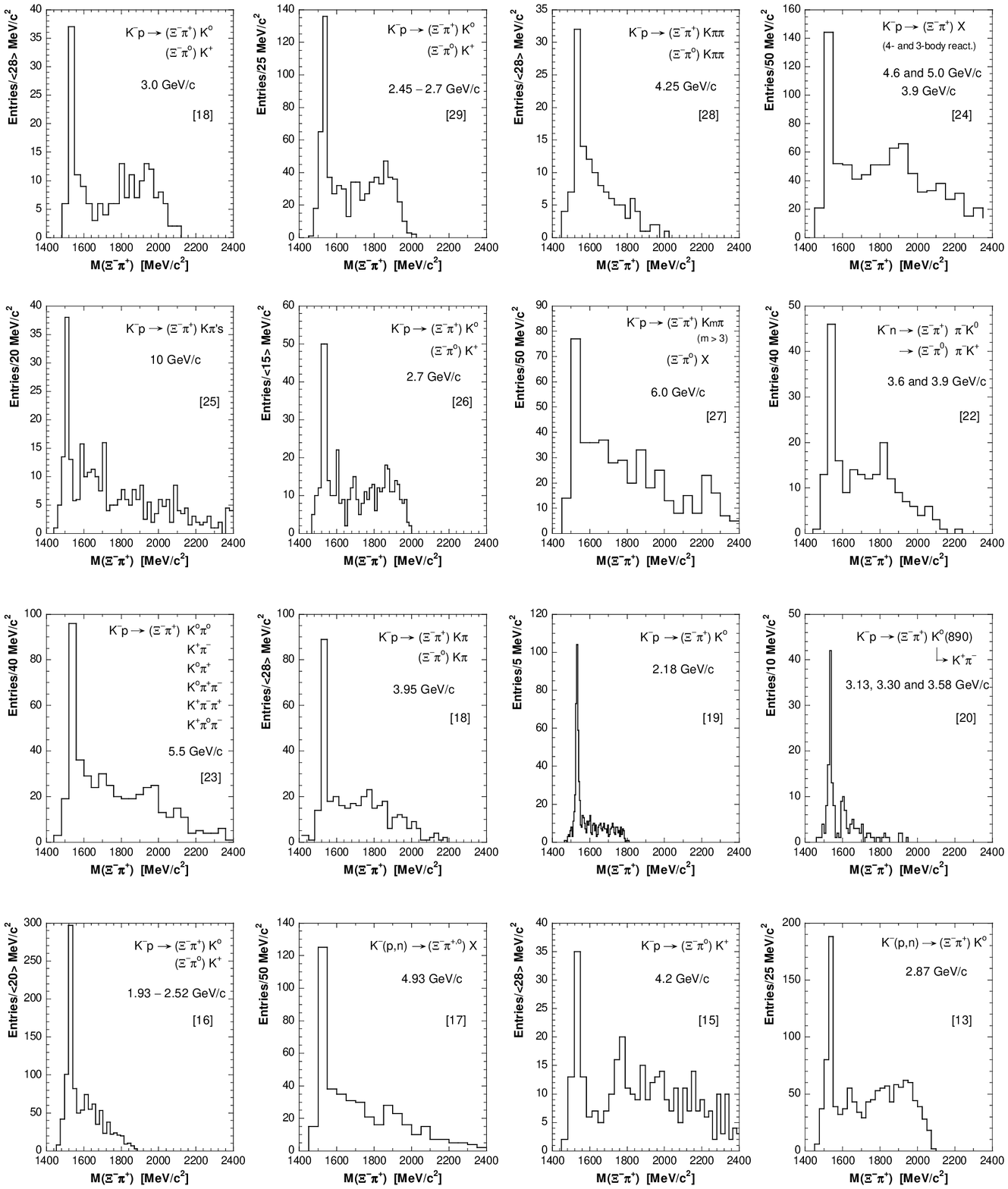,width=16cm}
\caption{Compilation of data on $\Xi^-\pi^{+0}$-spectroscopy.
}
\label{f1}
\end{figure}

The combinations $\Xi^-\pi^+$ and $\Xi^-\pi^0$ represent not only the $I_3=\pm1/2$ 
members of the pentaquark quadruplet but also the non-exotic isospin-doublet 
of excited $\Xi$ states.
A series of mass spectra from different experiments in
the lower range of event statistics is shown in Fig.~\ref{f1} in order to give
an overview of the situation without detailed individual discussion
which may be found in the original papers. The quadratic mass scale
used in some of the publications has been converted into a linear
one for ease of comparison, with the average linear bin width in the
measured region given on the vertical axis.

\vspace{2mm}
\noindent
Some comments are in place here.
\vspace{2mm}
\begin{itemize}
\item
All experiments see the lowest-lying $\Xi^*$ state at 1530 MeV/c$^2$
with between about 30 to 300 entries. 
This resonance can be used as gauge for the comparison of the sensitivity of the
different data sets, keeping in mind that the total number of $\Xi^{*0}$(1530) 
seen by NA49 is about 150 \cite{com}.
\item
There might be some indications for structure in the higher mass range, but all of 
them have low statistical significance.
\item
Cautious claims for definite mass states have been made in the mass ranges 1630, 
1820, and 1940~MeV/c$^2$. 
No indication of a structure in the 1860~MeV/c$^2$ region has been given by any author.
\end{itemize}
\vspace{2mm}

\begin{figure}
\centering
\epsfig{file=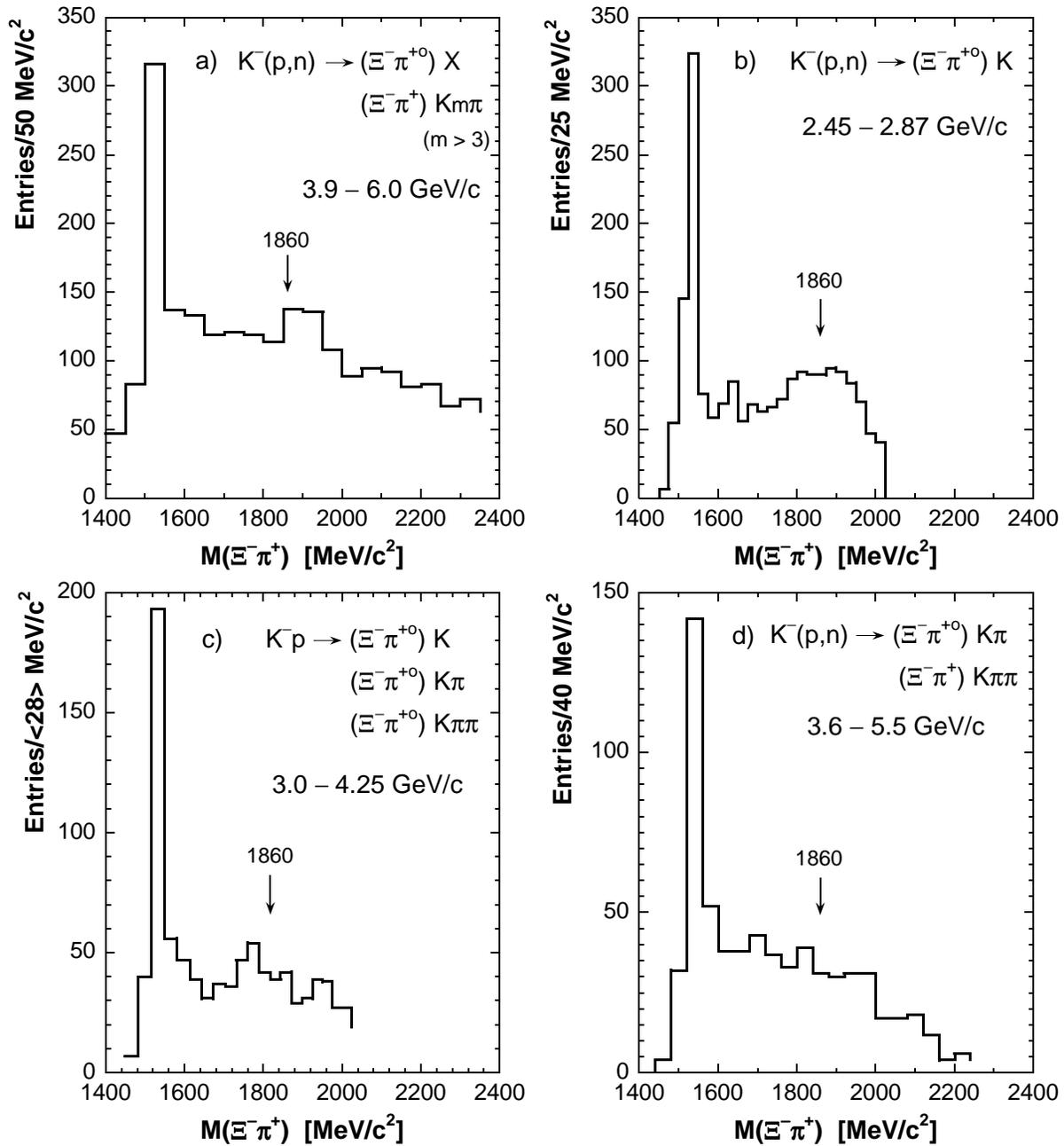,width=16cm}
\caption{Summary of data on $\Xi^-\pi^{+0}$-spectroscopy a) [24,17,27], b) [13,29], 
c) [28,18,15], d) [22,23].
}
\label{f2}
\end{figure}

In the few cases of identical binning, mass spectra can be added up
to improve on statistical significance. These summed mass distributions
are shown in Fig.~\ref{f2}. It is evident that in all cases the sensitivity
of NA49 is at least reached if not exceeded without showing any evidence for a 
signal at 1860~MeV/c$^2$. 

\begin{figure}
\centering
\epsfig{file=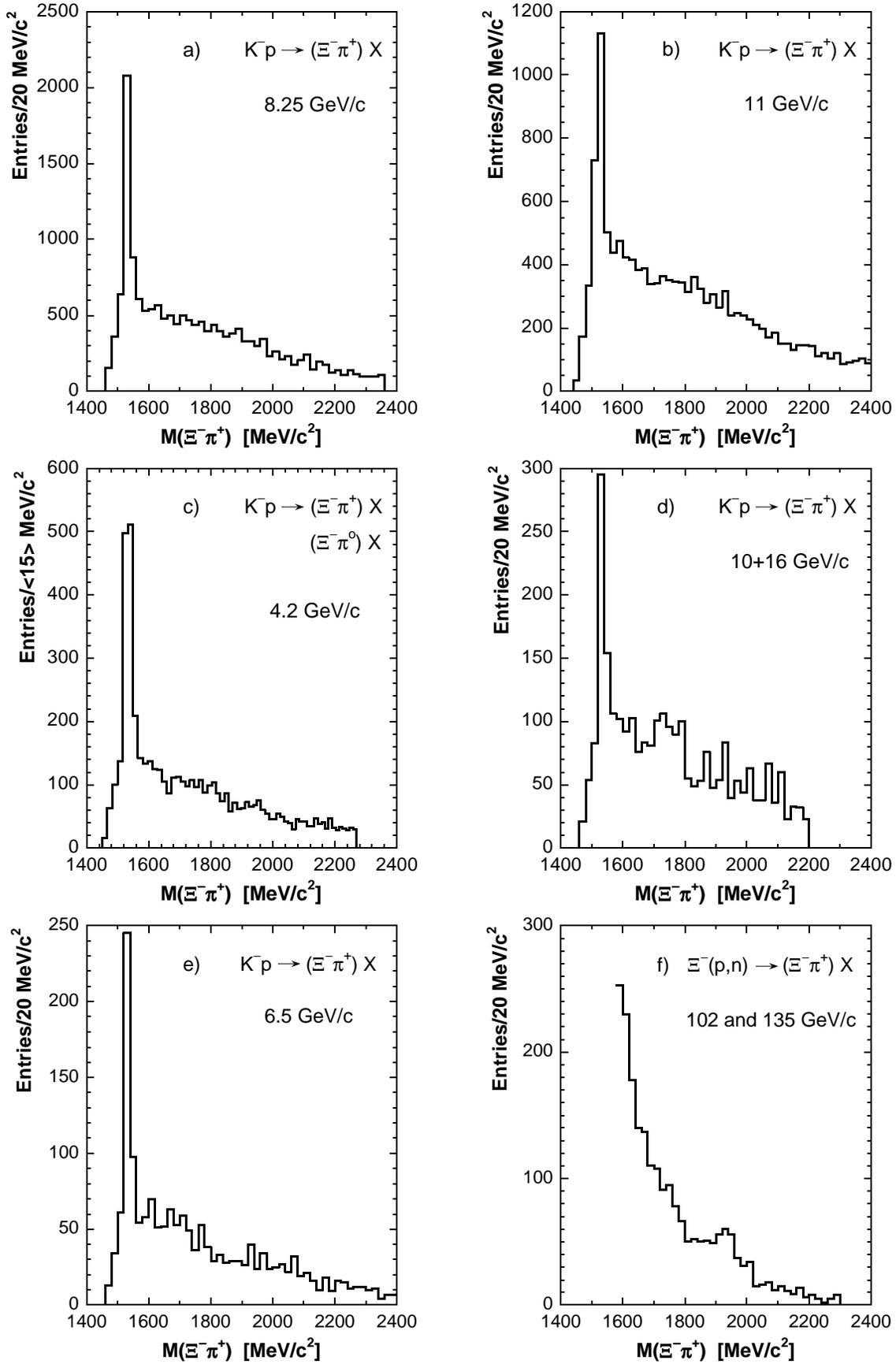,width=16cm}
\caption{Compilation of high statistics data on $\Xi^-\pi^{+0}$-spectroscopy
a) [9], b)[8], c) [14], d) [12], e) [11], f) [10].
}
\label{f3}
\end{figure}

A series of mass distributions from experiments with considerably higher
statistics is presented in Fig.~\ref{f3}. 
Here the number of entries in the $\Xi^{*0}$(1530) peak exceeds the NA49 data by 
factors of 2 to 15, with bin widths of 20 MeV/c$^2$ in most cases.
No structure in the region 1860~MeV/c$^2$ is visible in any of the distributions. 
In fact, the only claim for possible excited states comes from a hyperon
beam experiment (Fig.~\ref{f3}f) \cite{bia} where a broad structure around 1940~MeV/c$^2$ is
claimed to probably consist of several close sub-states.
A statistical analysis in this region with respect to a polynomial
background fit reveals a significance of order 4~$\sigma$ with no
indication for a signal at 1860~MeV/c$^2$.

\begin{figure}
\centering
\epsfig{file=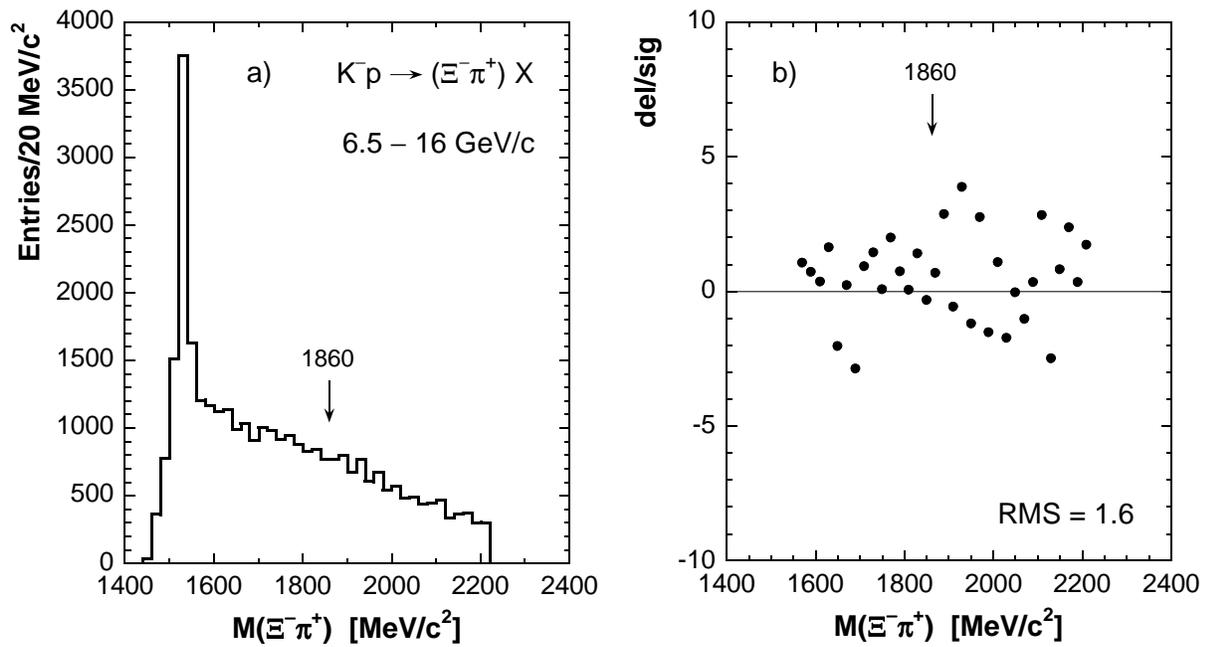,width=16cm}
\caption{a) Summary of high statistics data on $\Xi^-\pi^+$-spectroscopy [8,9,11,12]; 
b) deviation from
a straight line fit to the data (excluding the $\Xi^*$(1530)) in units of sigma.
}
\label{f4}
\end{figure}

Again, the mass distributions with equal binning can be added up,
yielding about 4500 entries in the $\Xi^{*0}$(1530) peak.
This sum distribution is presented in Fig.~\ref{f4}a together with a
statistical analysis of the mass region $1550-2200$ MeV/c$^2$ using
a polynomial fit (Fig.~\ref{f4}b). 
There is no evidence for a structure in this whole mass range beyond 2~$\sigma$, 
with a projected rms of 1.6.

Calibrating again with the $\Xi^{*0}$(1530) we conclude that the 
sensitivity of these combined data exceeds the one of the NA49
experiment by a factor of about 6, also taking into account
the lower combinatorial background at the lower energies. 
This means that a 2~$\sigma$ signal in the NA49 data should appear as
a 12~$\sigma$ peak in the combined data, which is clearly not visible
in the experimental residual distribution shown in Fig.~\ref{f4}b.

\begin{figure}
\centering
\epsfig{file=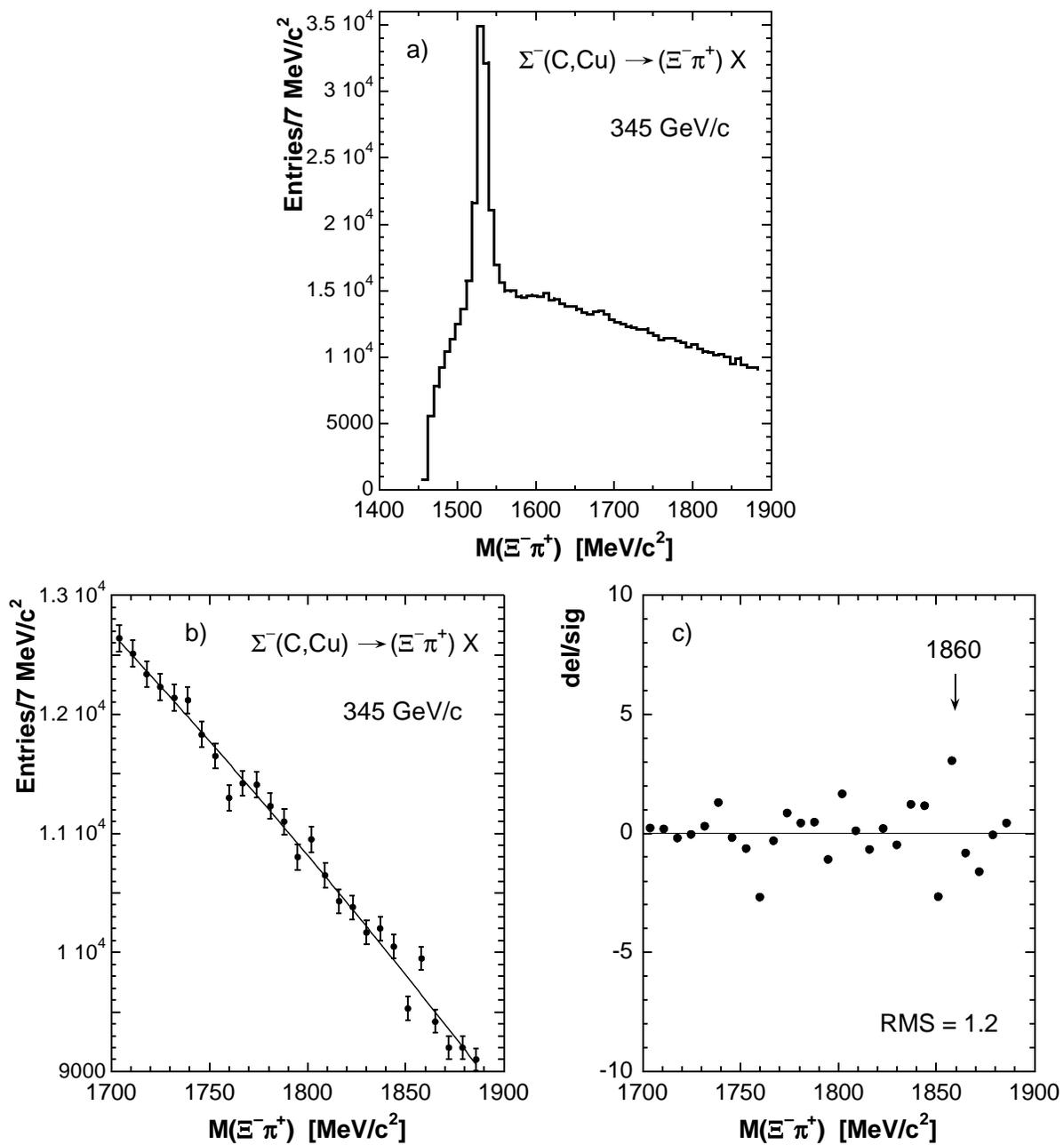,width=16cm}
\caption{a) Very high statistics $\Xi^-\pi^+$ invariant mass distribution \cite{adam};
b) polynomial fit to the data in the mass region 1700 -- 1900 ~MeV/c$^2$;
c) deviation from the polynomial fit in units of sigma.
}
\label{f5}
\end{figure}

Finally there are results from the hyperon beam experiment WA89 \cite{adam}
using the Omega Spectrometer at CERN. From a total of $10^8$~events this experiment produced
more than $2\cdot10^5$~$\Xi^-$ and about $6\cdot10^4$~$\Xi^{*0}$(1530)
leading to a sensitivity which exceeds the one of the NA49
experiment by a factor of 20.
The corresponding $\Xi^-\pi^+$ invariant mass distribution is presented in Fig.~\ref{f5}a, 
demonstrating the superior statistical accuracy of these data with a bin
width of 7~MeV/c$^2$.
The authors claim to observe the $\Xi^*$(1690) in this
channel, with a significance of only about 4~$\sigma$.
This may shed some light on the kind of event sample needed before
coming up with evidence for new states in this field.
The credibility of the signal is however enhanced in this case
by the fact that signals at the same mass have been seen before 
in different decay particle configurations. The mass region above 
1.7~GeV/c$^2$ is investigated by fitting the data with a polynomial (Fig.~\ref{f5}b)
and enumerating the bin-by-bin residuals (Fig.~\ref{f5}c). No structure is visible
at 1860 MeV/c$^2$ beyond $\pm3\sigma$.
The claim of the NA49 collaboration would, on the other hand, correspond to 
about 10~000 entries at this mass with a significance of more than 40~$\sigma$.

\subsection{The $\Xi^{*0}(1530)\pi^-$ Channel
}

\begin{figure}
\centering
\epsfig{file=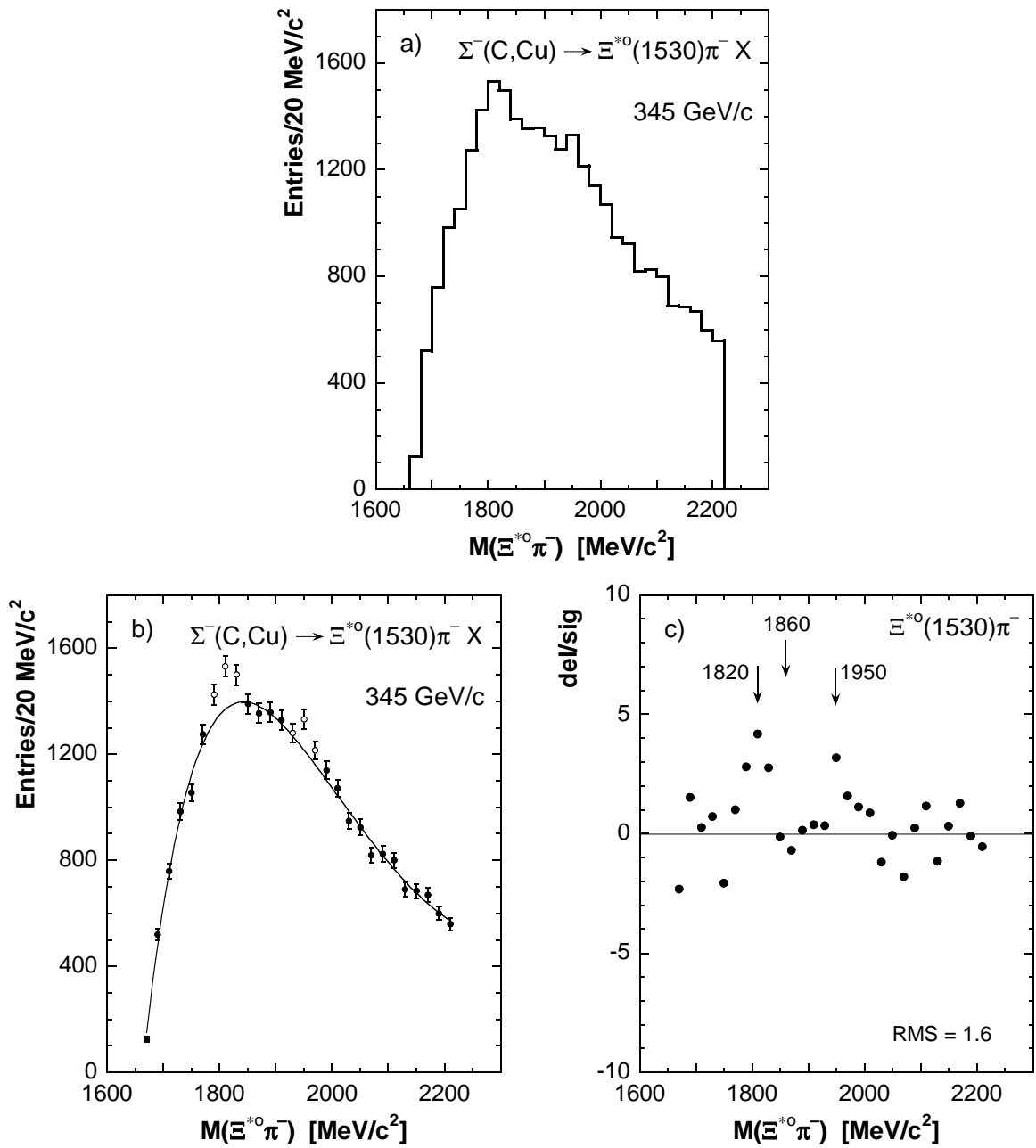,width=16cm}
\caption{a) $\Xi^{*0}(1530)\pi^-$ invariant mass distribution \cite{adam1};
b) polynomial fit to the data excluding the mass regions around 1820 and 1950~MeV/c$^2$
(open symbols);
c) deviation from the polynomial fit in units of sigma.
}
\label{f-ad1}
\end{figure}

The hyperon beam experiment \cite{adam1} also allows a look at the  $I_3=-1/2$
channel via the $\Xi^{*0}(1530)\pi^-$ mass combination. 
Here, two further resonances at 1820 MeV/c$^2$ and 1950 MeV/c$^2$ become visible with 
low statistical significance as shown in Fig.~\ref{f-ad1}. 
This indicates in the mass range above 1800 MeV/c$^2$ a preference for a cascade
decay via intermediate resonance states rather than a direct decay into two-body
ground state combinations. 
Again, no indication of a structure at 1860 MeV/c$^2$ 
is visible and again the high sensitivity needed to claim signals
beyond the non-exotic sector is exemplified.

\subsection{The $\Xi^{*0}(1530)\pi^+$ Channel
}

\begin{figure}
\centering
\epsfig{file=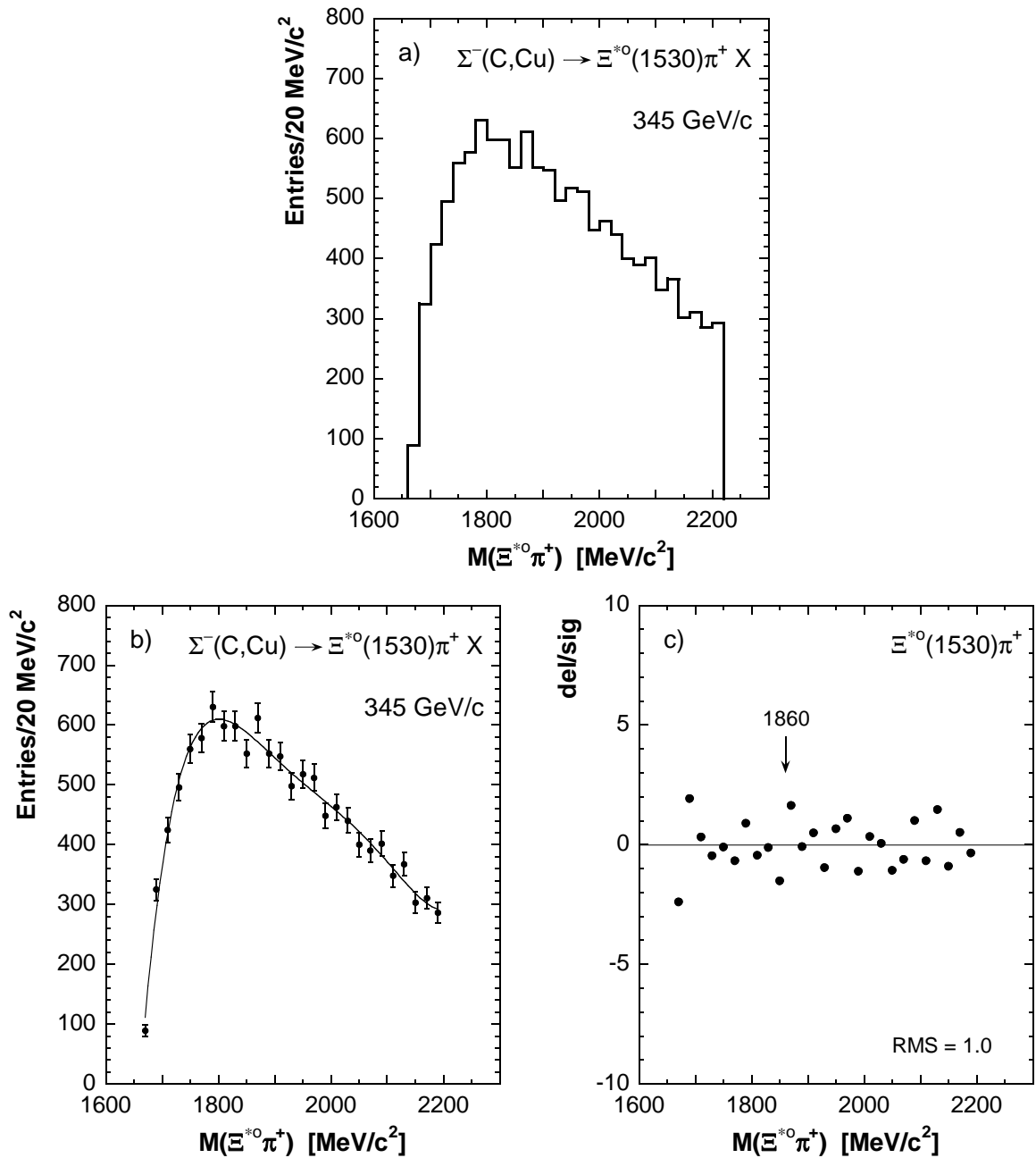,width=16cm}
\caption{a) $\Xi^{*0}(1530)\pi^+$ invariant mass distribution \cite{adam1};
b) polynomial fit to the data;
c) deviation from the polynomial fit in units of sigma.
}
\label{f-ad2}
\end{figure}

The exotic combination $\Xi^{*0}(1530)\pi^+$ corresponds to the $I_3=+3/2$
member of the pentaquark quadruplet. The respective invariant mass distribution
is also available in \cite{adam1} and is reproduced in Fig.~\ref{f-ad2}. 
A statistical analysis of this distribution over the
full mass range from 1700 to 2200 MeV/c$^2$ reveals no structure beyond
2~$\sigma$. 
It should be noted that also here the experimental sensitivity is more than 
one order of magnitude above the one of the NA49 experiment.

\subsection{The $\Xi^-\pi^-$ Channel
}
The interest in this exotic state was vivid mostly in the early days 
of $\Xi$ spectroscopy. The later, high statistics experiments generally
do not show the corresponding mass distributions, although in some
publications \cite{dib} \cite{ali} \cite{smi} the absence of structure in this 
channel is explicitly mentioned.

\begin{figure}
\centering
\epsfig{file=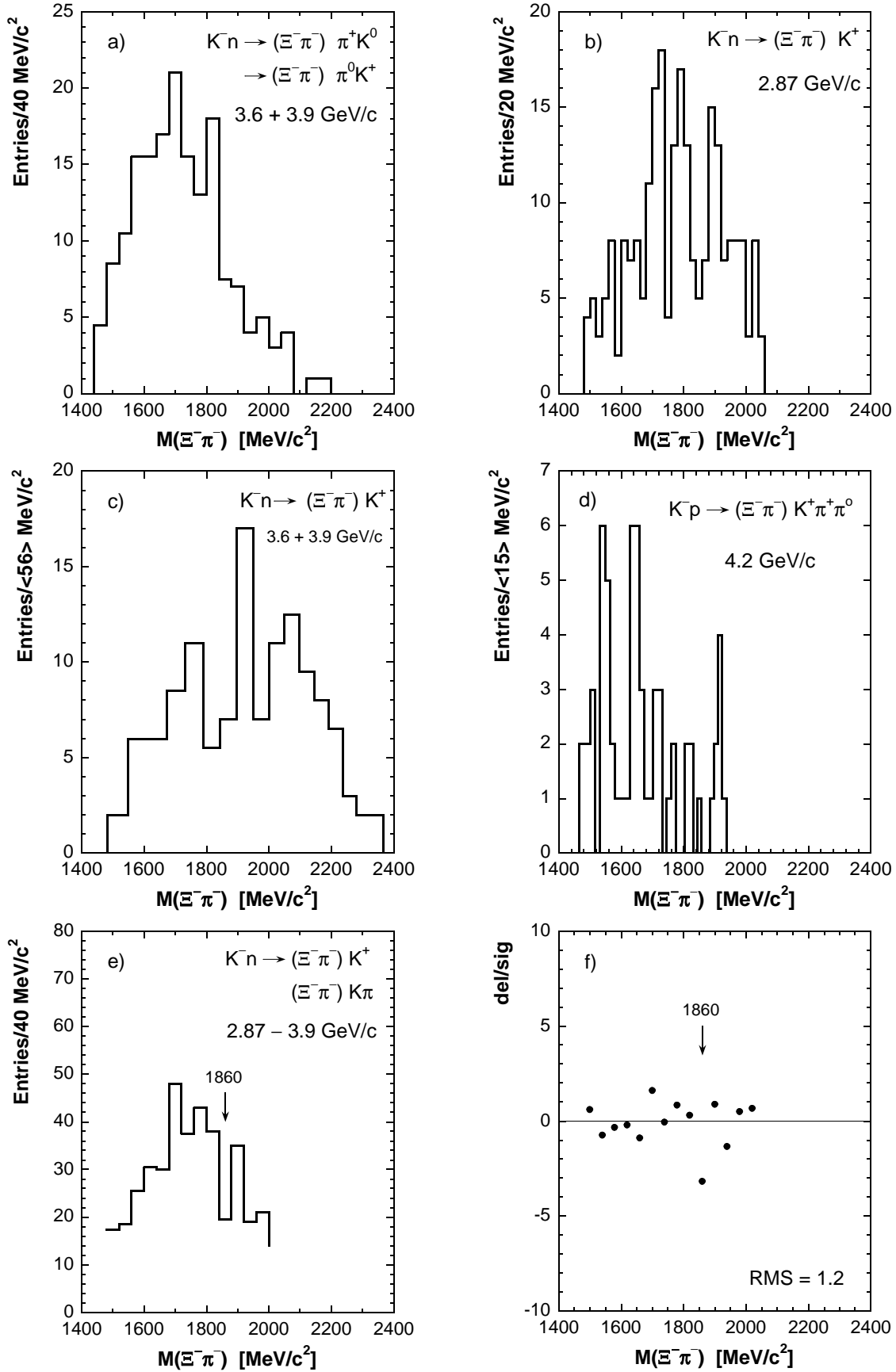,width=16cm}
\caption{Compilation of data on $\Xi^-\pi^-$-spectroscopy a) [22] b) [13] c) [22] d) [18];
e) sum of a) and b); f) deviation from a polynomial fit to e) in units of sigma.
}
\label{f6}
\end{figure}

We show in Fig.~\ref{f6}(a-d) the available invariant mass distributions from 4 publications. 
Although there are some uncertainties concerning the normalization of the respective 
final states, we estimate from similar distributions of the non-exotic state (see Fig.~\ref{f1}) 
that their event statistics corresponds to between 50 and 100 $\Xi^{*0}$(1530),
i.e. only a factor of about 1.5 to 3 below the NA49 sample. 
The first two distributions may be added up by doubling the binning in the first
one. This sum distribution is shown in Fig.~\ref{f6}e and the deviation from a polynomial 
fit in units of $\sigma$ in Fig.~\ref{f6}f. 
Again no structure is seen around 1860 MeV/c$^2$ with a sensitivity close to the one 
of NA49, but of course the bigger bin size has here to be considered.

Serious upper limits could be expected from the high statistics data 
\cite{bau} \cite{ast} \cite{gan} \cite{bia} \cite{adam} 
had they shown the spectra. The absence of 
this information does of course not imply that the according mass 
distributions have not been looked at. Nothing should therefore prevent 
the claimants of a new state to enquire about this with eventually 
still active members of the corresponding collaborations. Such a quest 
could be most successful with the hyperon beam experiment where data 
are still being published \cite{adam2}. In this case, the expected signal
of 20~000 entries in the peak with a significance in excess of 60~$\sigma$
(scaled from the NA49 claim) should indeed be rather difficult to miss. 

\section{ Internal Consistency
}

The NA49 experiment offers additional possibilities to study exotic
spectroscopy due to its wide acceptance including charged particle 
identification and due to the detection of neutrons in its hadronic 
calorimeter. 
The channels to be discussed here are pK$^+$ and nK$^+$.
They have recently been presented in a Letter of Intent submitted
to the SPSC Committee \cite{sps}. 
We will concentrate on an estimate of yields in relation to the $\Lambda$(1520) 
resonance and in relation to the claimed double-strange pentaquarks.

\subsection{ The $\Lambda$(1520) Resonance as Gauge Channel
}

\begin{figure}
\centering
\epsfig{file=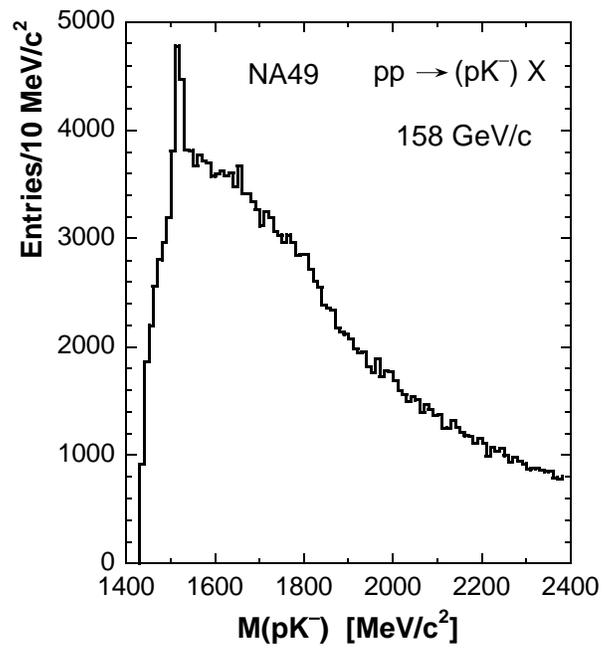,width=16cm}
\caption{pK$^-$ invariant mass distribution measured by NA49.
}
\label{f7}
\end{figure}

Fig.~\ref{f7} presents the pK$^-$ invariant mass distribution from NA49 using
the available event statistics in p+p interactions. 
About 4000 $\Lambda$(1520) hyperons are contained in the prominent peak
of this distribution. 
A more detailed analysis of this mass distribution would have to take account of the 
higher resonances above 1600~MeV/c$^2$ \cite{sps} but for the present discussion a 
rough estimate of the $\Lambda$(1520) yield is sufficient.
Including branching ratio, acceptance and fiducial cuts
this corresponds to about $4\cdot10^4$ produced $\Lambda$(1520) hyperons.

\subsection{The pK$^+$ Channel
}

\begin{figure}
\centering
\epsfig{file=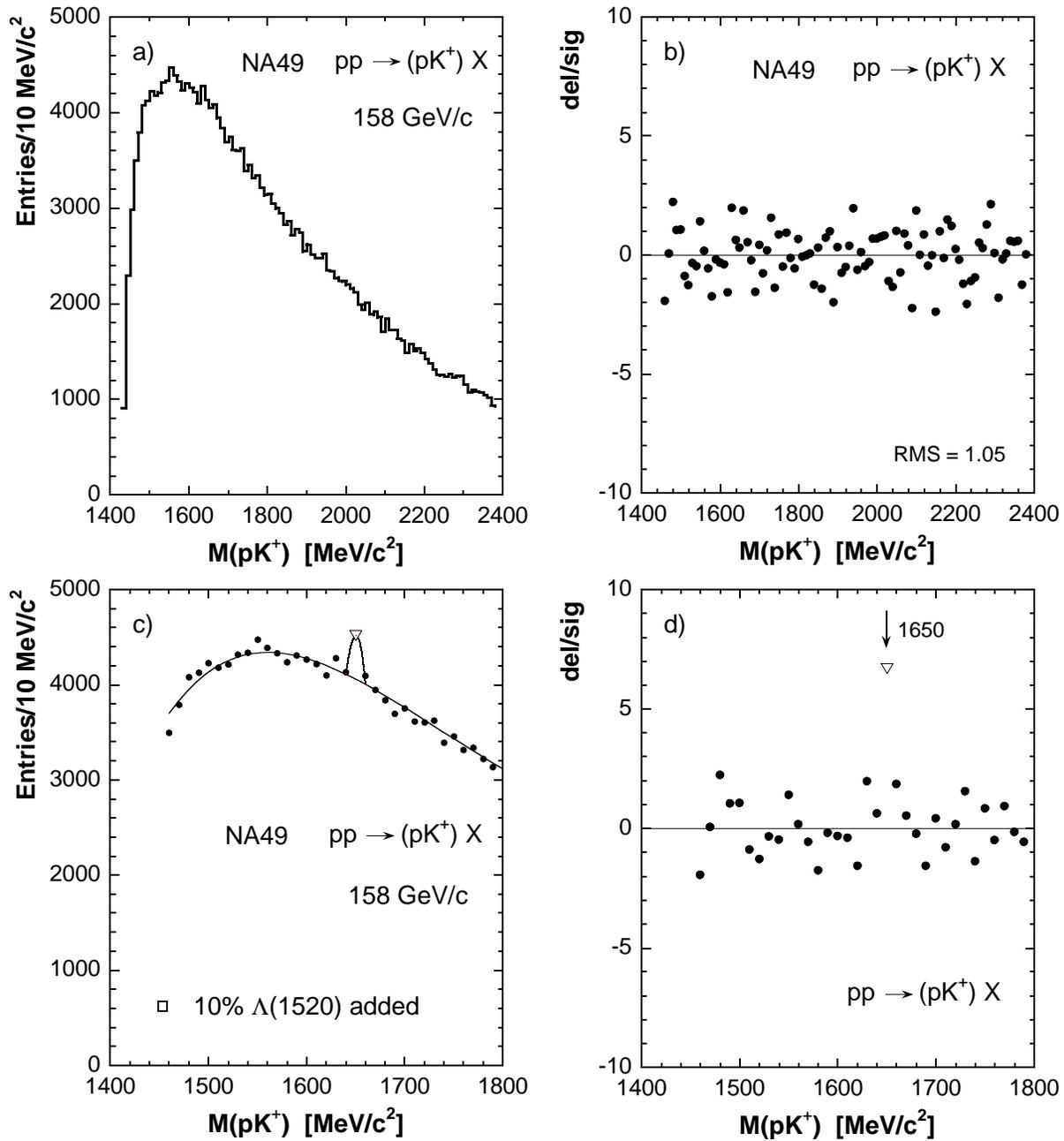,width=16cm}
\caption{a) pK$^+$ invariant mass distribution measured by NA49;
b)deviation from a multi-polynomial fit to the data in units of sigma;  
c) Isovector Pentaquark added to the pK$^+$ distribution at M$=1650$~MeV/c$^2$ (see text) 
assuming a yield of 10\% of the $\Lambda$(1520);
d) statistical significance of the added Isovector Pentaquark.
}
\label{f8}
\end{figure}

Fig.~\ref{f8}a shows the pK$^+$ invariant mass distribution for the same event sample,
again already presented in \cite{sps}. 
The statistical analysis of this mass distribution with respect to a multi-polynomial 
fit reveals no fluctuations beyond $2\sigma$, with a projected
rms of 1.05, over the full mass range (Fig.~\ref{f8}b). 
The corresponding pentaquark state (uudu$\overline{s}$) would be an $I=1$ partner 
of the $\Theta^+$(1540).
Assuming its mass to be 1650 MeV/c$^2$ \cite{wal}, a width of 10 MeV/c$^2$, and a yield of 10\% 
with respect to the $\Lambda$(1520), a signal as shown in Fig.~\ref{f8}c  would result, 
corresponding to a statistical significance of 7~$\sigma$ (Fig.~\ref{f8}d). 
We conclude that such candidates can be excluded over the full mass range on a level 
of 5\% with respect to the $\Lambda$(1520) yield.

\subsection{The nK$^+$ Channel
}

\begin{figure}
\centering
\epsfig{file=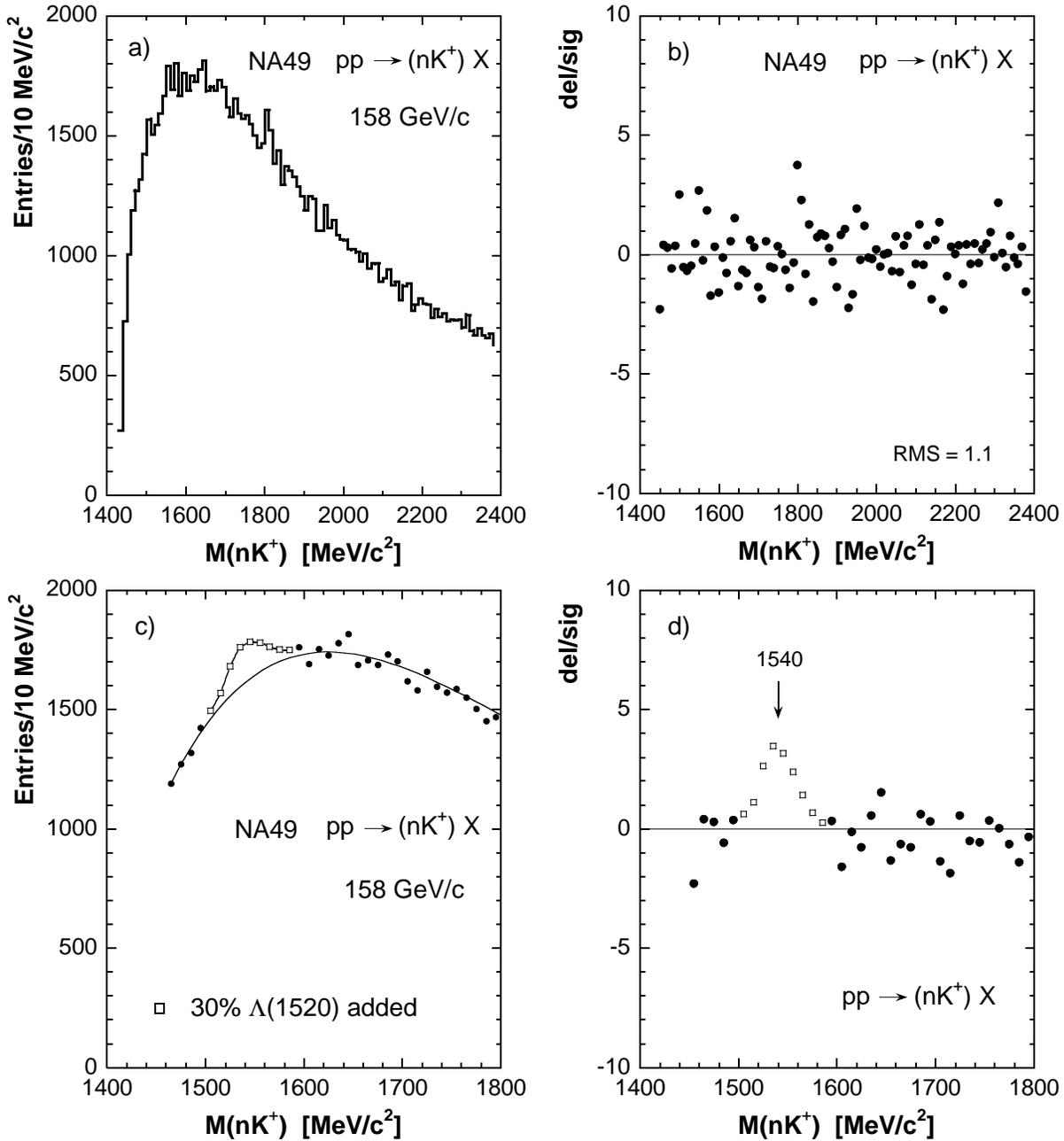,width=16cm}
\caption{a) nK$^+$ invariant mass distribution measured by NA49;
b)deviation from a multi-polynomial fit to the data in units of sigma;   
c) $\Theta^+$(1540) added to the nK$^+$ distribution at M$=1540$~MeV/c$^2$ assuming 
a yield of 30\% of the $\Lambda$(1520);
d) statistical significance of the added $\Theta^+$(1540).
}
\label{f9}
\end{figure}

The nK$^+$ invariant mass distribution is shown in Fig.~\ref{f9}a. 
Again with respect to a multi-polynomial fit, no deviation above the 2~$\sigma$ level 
is observed in the range around 1540 MeV/c$^2$ (Fig.~\ref{f9}b).
As argued in \cite{sps} this channel suffers in mass resolution due to
the performance of the NA49 hadron calorimeter. The expected
resolution in the 1540~MeV/c$^2$ range has been studied and evaluated
to a FWHM of 40~MeV/c$^2$ (see \cite{sps} for details).
Assuming the yield of a possible $\Theta^+$ resonance to be 30\% of the
$\Lambda$(1520), and convoluting the experimentally claimed width of the
$\Theta^+$ with the (dominant) experimental mass resolution, a predicted
"signal" as shown in Fig.~\ref{f9}c is produced. 
This would correspond to a 3-4~$\sigma$ deviation with respect to a smooth background 
over several bins (Fig.~\ref{f9}d) forming a 
shoulder which should be readily visible experimentally. 
We therefore conclude on a $\Theta^+$ yield of less than 30\%
of the $\Lambda$(1520) yield.

\subsection{Consistency Arguments
}

Even allowing for a rather wide margin of liberty in predicting the
relative yields of pentaquark states in the anti-decuplet for a
given energy and type of interaction, there are some basic considerations
that allow at least a very rough estimation of the relation between
the $S=+1$ singlet and the $S=-2$, $I_3=-3/2$ baryons.

\vspace{2mm}
\begin{itemize}
\item[a)]Reconstruction efficiency:  \\
The mean reconstruction efficiency for $\Xi$ hyperons in NA49 is known
to be 25\% \cite{barna}.

\item[b)] Additional cuts:  \\
The acceptance for $\pi$ and the additional cuts introduced results in an effective loss
of at least a factor of 2 for the $\Xi^-\pi^-$ combination.

\item[c)] Branching fraction:  \\
In addition to the  $\Xi^-\pi^-$ decay also the channels Y$\overline{\mbox{K}}$, 
$\Xi^*(1530)\pi$ and other multi-pion final states are open. 
We therefore estimate the two-body $\Xi^-\pi^-$ branching fraction to less than 20\%.

\item[d)] Charge conservation:  \\
In hadronic collisions there is in general a charge penalty to be
payed if produced particles deviate far from the initial charge
configuration. For pp interactions, this factor should be at least
4 between a positive and a double-negative state. 

\item[e)] Strangeness conservation:  \\
Similar penalty factors exist for each step in strangeness content
of a produced particle with respect to the initial state. Very
conservatively this factor can be estimated to about $10-20$ between
the two states concerned.
\end{itemize}
\vspace{2mm}

\noindent
Contracting these factors and starting from the number of entries
in the $\Xi^-\pi^-$ peak, we arrive at more than 80~000
$\Theta^+$ baryons to be produced in the NA49 experiment. This yield,
which has to be regarded as a lower limit, exceeds by far the
known production rate of $\Lambda$(1520) hyperons. This is inconsistent
with the upper limit of 30\% of $\Lambda$(1520) established in Section 3.3
above.

This consistency argument stresses the necessity of keeping track
of relative yields and production cross sections in all stages of the
experimental studies.

\section{Past Claims Concerning Exotic or Rare Hadron Spectroscopy 
}

In the present context it might be useful to look back on two
historic examples of claims which suffered from insufficient statistical
accuracy.

35 years ago, a spectrometer experiment announced the discovery of the
deviation of the a$_2$(1320) resonance from the standard Breit-Wigner
form to a dipole-type shape with a significance of 6~$\sigma$. 
No less than five further, different experiments found
similar deviations over the following two years, with significances
in the range of 3~$\sigma$. It took more than four years 
of work and the results of a superior BNL experiment to disclaim the 
split with a more than 20~$\sigma$ significance in 1971. See the review 
of R.H.Dalitz \cite{dal} for an excellent description of the facts and the 
statistical problems involved.

25 years ago, and after the masses of heavy flavour hadrons had become 
known from work at e$^+$e$^-$ storage rings, there was a host of claims for 
hadronic heavy flavour production by different experiments and 
collaborations at the CERN-ISR. These claims started with the D$^+$ production 
\cite{dri1} in 1979 and culminated with experimental evidence for beauty 
baryon production \cite{bas} in 1981. 
A common feature of all these claims 
was a statistical significance of typically 4-6~$\sigma$
with a low number of some dozen entries in the corresponding peaks.
This is very reminiscent of the present situation around the different
pentaquark claims.
  
When finally serious cross sections were elaborated \cite{dri2} and compared
to strict upper limits from lepton pair production \cite{hgf}, the majority
of the claims turned out to be above those limits by typically 1 to several
orders of magnitude.

\section{Conclusion
}

Following the claim of $S=-2$ pentaquark states by the NA49 collaboration,
published mass distributions in the $\Xi\pi$ and $\Xi^*(1530)\pi$ channels
have been re-investigated. As a result, for three of the four $S=-2$ states
in the minimal pentaquark anti-decuplet, signals at the quoted mass
of 1860 MeV/c$^2$ can be excluded with a sensitivity which is at least
one order of magnitude above the one of the NA49 data.

Exclusion on the same level is, for the time being, not possible for
the $I_3=-3/2$, $\Xi^-\pi^-$ channel, because the mass spectra have not been 
published for the corresponding high-statistics experiments. Here,
the re-establishment of the mass distributions from the existing data
is advocated.

It is, however, argued that the claimed number of entries in this channel
is inconsistent with the absence of any signal in the nK$^+$ decay
of the $\Theta^+$(1540) state studied with the same NA49 data set.

In this context, some of the negative experiences accumulated in the
past with low-statistics, low-significance hadron spectroscopy, are
recalled.  

%--------------------------------------------------------------------------------
%\newpage

\clearpage
%-----------------------------------------------------------------------------

\clearpage

\clearpage

\clearpage

\clearpage

\clearpage

\clearpage

\clearpage

\clearpage

\clearpage

\clearpage

\clearpage

%-----------------------------------------------------------------------------


\begin{thebibliography}{[20]}
\bibitem{com} i.e. roughly 10\% of the total number of observed $\Xi^-$ 
which is about 1600 \cite{pen}.
\end{thebibliography}
\end{document}